\documentclass{iopconfser}
\usepackage[numbers,sort&compress]{natbib}
\usepackage[dvipsnames]{xcolor}
\usepackage{xspace,slashed}
\usepackage{graphicx}
\usepackage{amssymb}

\begin{document}
\newcommand{\comment}[1]{{\color{persianred}\texttt{NOTE:} #1}}
\newcommand{\Qibolab}{\texttt{Qibolab}\xspace}
\newcommand{\Qibo}{\texttt{Qibo}\xspace}
\newcommand{\Qibocal}{\texttt{Qibocal}\xspace}
\newcommand{\Qibosoq}{\texttt{Qibosoq}\xspace}
\newcommand{\cell}[2][c]{\begin{tabular}[#1]{@{}c@{}}#2\end{tabular}}

\vspace*{-1.5cm}
\begin{flushright}
CERN-TH-2024-126\\
\end{flushright}
\vspace{0.3cm}

\title{ An open-source framework for quantum hardware control }

\author{Edoardo Pedicillo$^{1,2}$, Alessandro Candido$^{3}$, Stavros Efthymiou$^{1}$, 
Hayk Sargsyan$^{1}$, Yuanzheng Paul Tan$^{6}$, Juan Cereijo$^{1}$, Jun Yong Khoo$^{4}$, Andrea Pasquale$^{1,2}$,
Matteo Robbiati$^{2,5}$ and Stefano Carrazza$^{1,2,3}$}

\affil{$^1$Quantum Research Center, Technology Innovation Institute, Abu Dhabi, UAE}
\affil{$^2$TIF Lab, Dipartimento di Fisica, Università degli Studi di Milano}
\affil{$^3$Theoretical Physics Department, CERN, 1211 Geneva 23, Switzerland}
\affil{$^4$Institute of High Performance Computing (IHPC), Agency for Science, Technology and Research (A*STAR), 1 Fusionopolis Way, \#16-16 Connexis, Singapore 138632, Singapore}
\affil{$^5$European Organization for Nuclear Research (CERN), Geneva 1211, Switzerland}
\affil{$^6$Division of Physics and Applied Physics, School of Physical and Mathematical Sciences, Nanyang Technological University, 21 Nanyang Link, Singapore 637371, Singapore}

\begin{abstract}

    The development of quantum computers needs reliable quantum hardware 
    and tailored software for controlling electronics specific to various quantum platforms. 
    Middleware is a type of computer software program that aims to provide
    standardized software tools across the entire pipeline, 
    from high-level execution of quantum computing algorithms to low-level driver
    instructions tailored to specific experimental setups, including instruments.
    This paper presents updates to \Qibolab, a software library that leverages 
    \Qibo's capabilities to execute quantum algorithms on self-hosted quantum 
    hardware platforms. 
    \Qibolab offers an application programming interface (API) for instrument control
    through arbitrary pulses and driver operations including sweepers.
    This paper offers an overview of the new features implemented in \Qibolab since Ref.~\cite{Efthymiou_2024},
    including the redefined boundaries between platform and channel classes, the integration of an emulator 
    for simulating quantum hardware behaviour, and it shows updated execution times benchmarks
    for superconducting single qubit calibration routines.

\end{abstract}

\section{Introduction}

Nowadays, quantum computing research and development needs both reliable quantum hardware and classical hardware. 
The latter is what is usually called the control electronics and takes care 
of synthesizing the simultaneous synchronized control waveforms, specific to each different quantum
platform technology, required to operate quantum hardware.
The objective of middleware is to provide standardized software tools for the whole pipeline,
e.g., from the high-level execution of quantum computing algorithms based on the quantum circuit
paradigm, to operate the low-level control instrument instructions tailored to a specific experimental setup.

The commitment to build a software with a platform-agnostic interface helps the transition from theory
to experiments by reducing the effort and expertise required to operate a quantum platform
and develop novel quantum algorithms.
Additionally, it ensures reusability and this allows the creation of a community of quantum 
laboratories that can share experiments and data significantly reducing 
the burden coming from the different setups. 

Our commitment to open-source and community-driven software started with
\texttt{Qibo}~\cite{Efthymiou_2021}, a framework for gate-based and adiabatic 
quantum computing with hardware acceleration.

In this paper, we present updates to \Qibolab~\cite{Efthymiou_2024}, a software library 
that leverages \Qibo’s potential to execute quantum algorithms on self-hosted
quantum hardware platforms.

\Qibolab takes care of all the necessary operations to prepare the execution of quantum circuits
on a fully characterized device.

\Qibolab is running on a host computer, which communicates, typically via a
network protocol, with the control electronics used for generation of synchronized signal sequences. 
These electronics are connected to the quantum processing unit (QPU) via different channels. 

In the case of operating a superconducting quantum device: the readout and
feedback channels in a closed loop for measuring the qubit, the drive channels
for applying gates and, for flux-tunable qubits, the flux channels for tuning
their frequency.
Moreover, some architectures there may have coupler qubits between some 
computational qubits that need flux pulses to control computational qubit interactions.

The library offers a dedicated application programming interface (API) not only for quantum circuit design,
but also qubit calibration, instrument control through arbitrary pulses, driver operations 
as sweepers.

Thanks to \Qibolab, we can operate the control electronics required to fully calibrate a quantum device performing
specific experiments. In order to make this process as smooth as possible, we have also
developed on top of \Qibolab the \texttt{Qibocal}~\cite{Qibocal} library collecting some calibration
protocols for superconducting qubits with a user-friendly interface.

Successful implementation of the whole \Qibo ecosystem will provide the research community with a prototype
of an extensible, quantum hardware-agnostic, open-source hybrid quantum operating system, 
fully tested and benchmarked on superconducting platforms.

In the following sections, we first give an overview of the project in Sec.~\ref{sec:software},
highlighting the differences with the old software version described in Ref.~\cite{Efthymiou_2024}, 
then we provide the updated time benchmarks in Sec.~\ref{sec:benchmark}, at the end we
provide some details about the emulator in Sec.~\ref{sec:emulator}.

\section{Project Overview}
\label{sec:software}

\Qibolab provides four main interface objects:
the \texttt{Pulse} object, which is used to define arbitrary pulses played on the quantum device,
a series of pulses can be collected in \texttt{PulseSequence} and executed on 
a specific QPU through the \texttt{Channels}.

Pulses constitute the basic building blocks of programs executed on quantum hardware.
They represent a physical pulse, i.e., they are used to read 
the state of a qubit, drive it to change its state,
or flux bias a qubit to change its resonant frequency to probe two-qubit interactions.

\Qibolab provides pulse objects for each one of these operations as its
\texttt{Pulse} object holds the required information about amplitude,
duration and phase of the pulse for the generation of physical pulses.
Differently from the previous version, the \texttt{Pulse} object now further 
facilitates the integration with different control electronics.

Abstract pulse sequences defined using the \texttt{Pulse} API can be deployed on
hardware using a \texttt{Platform}. \texttt{Platform} is the \texttt{Qibolab} core object and it
orchestrates the different instruments for qubit control.
Each \texttt{Platform} instance corresponds to a specific quantum device controlled
by a specific set of instruments. It allows users to execute a single sequence,
a batch of sequences, or perform a sweep, in which one or more pulse parameters
are updated in real-time, within the control instrument without external communication. 
This new \Qibolab version allows the \texttt{Platform} to execute all of them
within a single instrument connection step, increasing the uptime of the QPU.

The \texttt{Platform} object also contains single qubit, two qubit interactions 
and coupler qubits information and eventually any other quantum components like 
parametric amplifiers that may be present in the experimental setup.
Since, computational qubits and coupling qubits are similar objects from the 
instrument point of view, as elements you send pulses to, so in the software level 
they can be defined by the same class we called \texttt{QuantumElement}.

To define two qubit interactions, \texttt{QubitPair} objects 
contain information about coupled qubits pairs in a given device for a given 
configuration and their corresponding two-qubit native gates.

Finally, \texttt{Channel} represents the connection between the quantum device 
inputs and outputs to the proper instrument port.This connection is essential 
for the instrument our interface to target the desired \texttt{QuantumElement}.
It also provides a \texttt{QuantumElement}-centric interface instrument parameter setting,
which is useful in calibration routines.

\texttt{Channels} are also responsible of signal generation. 
Differently from the previous \Qibolab version, now the readout is decoupled
into probe and acquisition channels, thus giving more freedom to the end user to decide
which part of the readout pulse to acquire or extend the acquisition beyond the 
time interval defined by the readout pulse.

With the new refactoring, the boundaries between the \texttt{Platform} and \texttt{Channel}
classes are redefined: there is no more distinction between readout 
and drive pulses. Since, they are just defined by the physical channel where
the pulses are sent, the frequency is now a \texttt{Channel} property, 
so this class is closer to a logical definition of channel.

Another significant change is \texttt{Transpiler} which got moved
to \Qibo.  This choice is driven by a rethinking of the respective roles
of the \Qibo ecosystem libraries and transparency. 
The transpilation procedure consists of processing the sequence of gates 
required by an algorithm and integrate SWAP gates, if necessary, so that the connectivity
of the chip is respected.
In practice, given the topology of the device, this is a task completely solvable within \Qibo itself.
Once a connectivity-compatible sequence of gates is defined, \Qibolab provides a compiler,
that is, the translation of abstract gates into native gates, and thus into pulses.

\begin{figure}
\centering
\includegraphics[width=.88\textwidth]{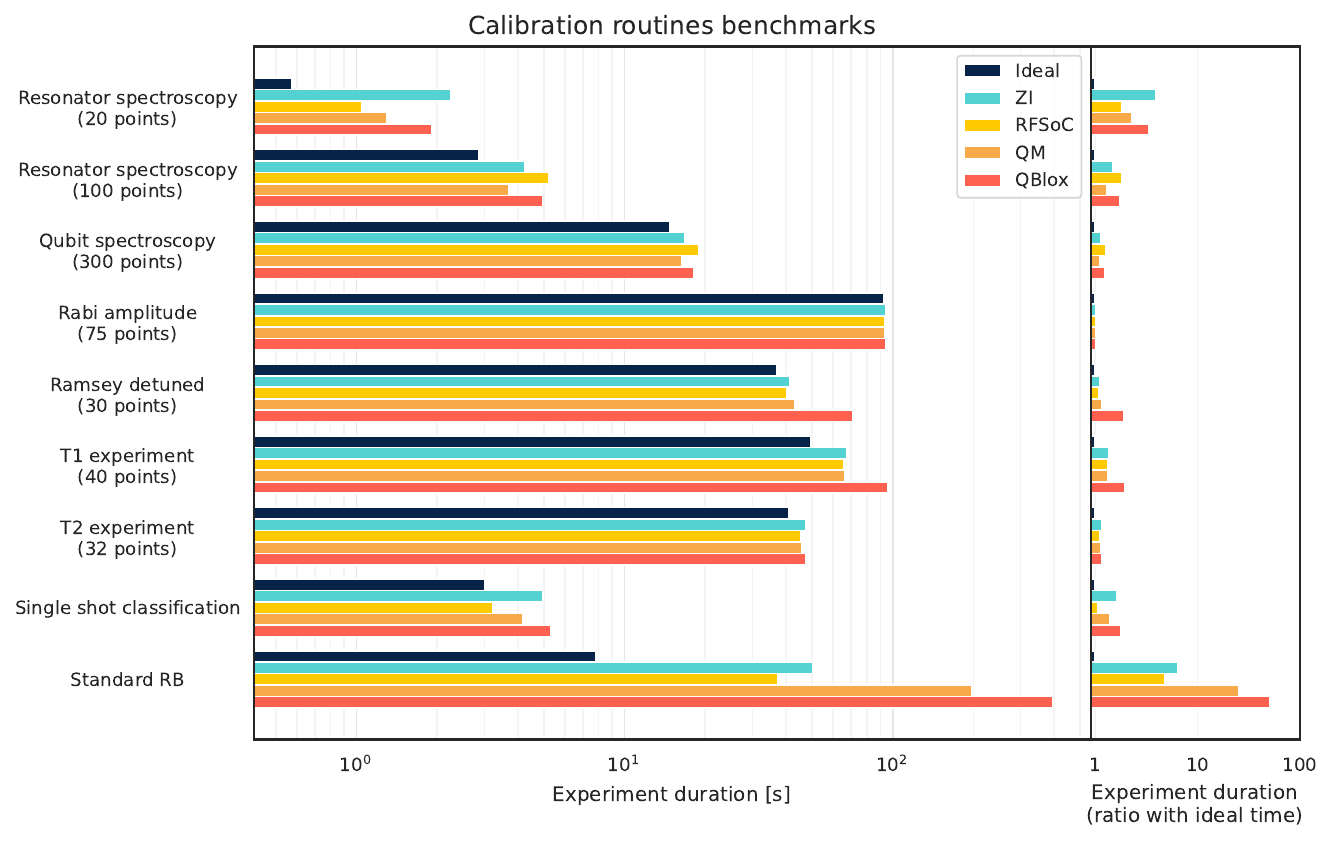}
\caption{Execution time of different qubit calibration routines on various 
    electronics as in Ref.~\cite{Efthymiou_2024}. On the left side, we show the absolute times in seconds 
    for each experiment. The ideal time (black bar) shows the minimum time 
    the qubit needs to be affected in each experiment. 
    On the right side, we calculate the ratio between the actual execution time and 
    the ideal time. Real-time sweepers are used, if supported by the control device, 
    in all cases except the \textit{Ramsey detuned} and \textit{Standard RB} experiments.
\label{fig:benchmark}}
\end{figure}

\section{Cross-platform benchmark}
\label{sec:benchmark}

In this section, we show the updated benchmarks already presented in the \Qibolab~paper.
They are related to the execution times of a set of calibration experiments performed
on a single qubit by different control electronics (see Tab.\ref{tab:drivers}) currently supported by \Qibolab~out of the box.

In contrast to the previous benchmark version, \Qibolab~side, we have implemented the possibility to unroll a list of
sequences to a single pulse sequence that contains multiple measurements. This approach achieves much faster execution 
compared to executing the sequences one by one in a software loop. On the side of the control electronics, we have updated the 
Zurich Instrument software to \texttt{LabOneQ 2.16.0}~\cite{LabOneQ}.

The experiments chosen for this benchmark represent the minimal set of routines 
required for superconducting single qubit calibration. 

They also offer a view of the different execution modes supported by \Qibolab: in particular 
the \textit{Single shot classification} experiment executes fixed pulses sequences, 
while the \textit{Spectroscopies} perform different sweeps over pulse parameters.

All the results are summarized in Fig.~\ref{fig:benchmark}, for each calibration protocol 
we provide also the theoretical execution time $\mathrm{T_{ideal}}$, 
defined as the total duration of the pulse sequences executed
during the acquisition, whose formula~\cite{Efthymiou_2024} is
\begin{equation}
    \mathrm{T_{ideal}} =n_{\mathrm{shots}} \sum_i (T_{\mathrm{sequence}, i} + T_\mathrm{relaxation}).
\end{equation}
where $T_{\mathrm{sequence}, i}$ is the duration of the whole pulse sequence
in the $i$-th point of the sweep, $T_{\mathrm{relaxation}}$ the time we wait
for the qubit to relax to its ground state between experiments,
$n_{\mathrm{shots}}$ the number of shots in each experiment and
the sum runs over all points in the sweep.
The $\mathrm{ideal}$ time denotes how long the qubit is used during
an experiment and provides the baseline for our benchmark.

As already pointed out in Ref.~\cite{Efthymiou_2024}, the comparison between
the ideal and real execution times show that the first one is always less than the second one, 
because of different overheads, indeed we can express the real execution time $\mathrm{T_{real}}$ as

\begin{equation}
  \mathrm{T_{real}} = T_{\mathrm{qibo}} + T_{\mathrm{inst}} + \mathrm{T_{ideal}},
\end{equation}
 
where $T_{\mathrm{qibo}}$ is the overhead coming from \Qibolab~backend and $T_{\mathrm{inst}}$ the instruments one.
We found that the overhead coming from the \Qibolab, $T_{\mathrm{qibo}}$, 
is negligible compared to that of the control instruments, i.e.,

\begin{equation}
    \mathrm{T_{real}}  \simeq T_{\mathrm{inst}} + \mathrm{T_{ideal}}.
 \end{equation}

\begin{table}
	\centering
    \begin{tabular}{lcc}
    \hline \hline
	    \textbf{Device}     & \textbf{Firmware}   & \textbf{Software}       \\ \hline
        Qblox               & 0.4.0               & qblox-instruments 0.9.0~\cite{Qblox} \\
        QM                  & QOP213              & qm-qua 1.1.1~\cite{qm_qua}            \\
        Zurich              & Latest (November 2023)  & LabOneQ 2.16.0~\cite{LabOneQ}          \\

        RFSoCs              & Qick 0.2.135~\cite{Stefanazzi2022}        & Qibosoq 0.0.3~\cite{Qibosoq}           \\ \hline
    	Erasynth++          & -                   & -                       \\
        \cell{R\&S SGS100A}     & -                   & QCoDeS 0.37.0~\cite{QCODESsite}           \\
   \hline \hline
   \end{tabular}
	\caption{Outline of the devices and firmware/software version supported during the benchmark.}
	\label{tab:drivers}
\end{table}

\section{Emulator}
\label{sec:emulator}

The new \Qibolab version supports emulators to simulate quantum hardware. It is a crucial tool especially for \Qibocal, as it
could be used both for testing the software itself (especially when access to the real device is limited or unavailable)
and serve as a digital twin to a specific quantum hardware to assist with a range 
of functionalities including predicting output of calibration experiments, 
coarse-grain calibrations, pulse-shaping and test-bedding, and more.
For the latter, the emulator requires the device parameters that characterize 
the quantum hardware of interest as inputs to build an accurate model of the hardware.

To integrate the emulator into the \Qibolab~ecosystem, we deployed an ad-hoc controller class called \texttt{PulseSimulator}.
The \texttt{PulseSimulator} is exclusively called by the emulator and it serves
primarily as a middleman to translate and communicate objects between \Qibolab~and 
the selected quantum dynamics simulation library (hereafter referred to as the
simulation engine) used to numerically solve the underlying quantum dynamics of
the device model in the presence of time-dependent control pulse sequences. 
Specifically, it initializes the simulation engine with a device model specified
by the device parameters and simulation settings in the runcard, extracts 
modulated signal waveforms from \Qibolab pulse sequences and sends them to 
the simulation engine which in turn performs the dynamics simulation, and at
the end of the simulation, translates the results generated from the simulation
engine back to  \Qibolab (or \Qibo) result objects.

\begin{figure}[h]
\centering
\includegraphics[width=.88\textwidth]{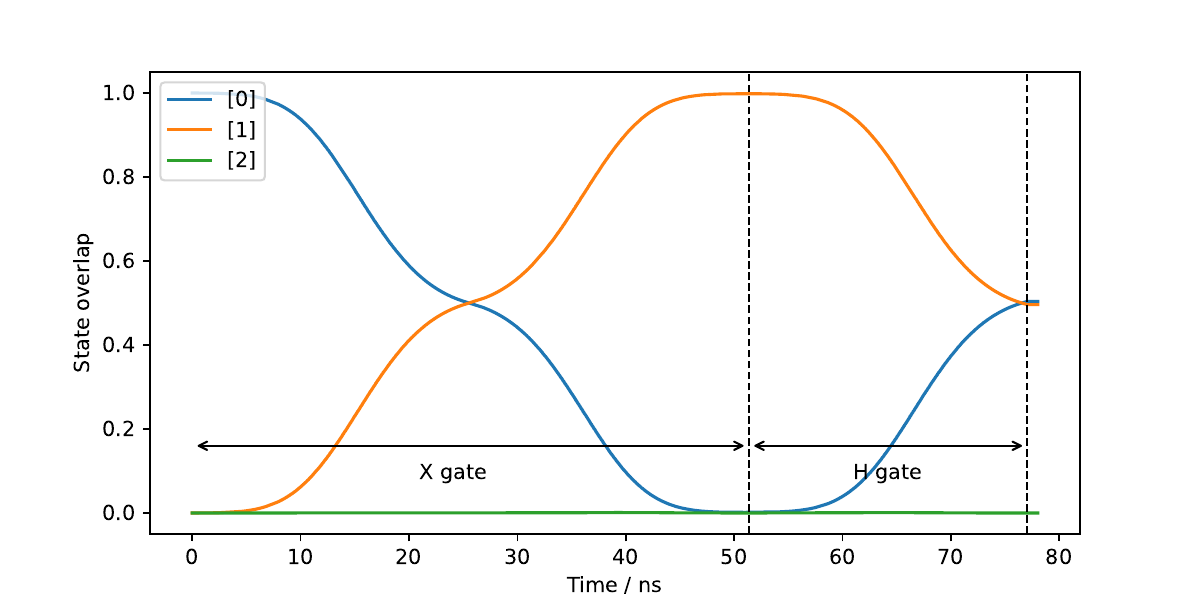}
\caption{State overlap between the simulated qubit modelled as a three-level 
    system with each of its energy modes as it evolves under a control pulse 
    sequence for an X gate followed by a Hadamard (H) gate.
\label{fig:sim_plot}}
\end{figure}

From the user's point of view therefore, the emulator platform behaves like 
a quantum hardware platform and both are equivalent in terms of attributes and 
functions. Consequently, it requires a platform folder and it is initialized 
as any other device platform. In addition, the emulator returns simulated 
results -- a time series of statevectors (or density matrices when including dissipation)
of the underlying quantum system that was generated sequentially by the simulation
engine as it solves the dynamics described by the Lindblad master equation,
\begin{equation}
\dot{\rho} (t) = -i [ H (t), \rho (t)] + \sum _{k} \frac{\gamma_k}{2} \left[ 2 A_k \rho(t) A_k^\dagger - \rho (t) A_k^\dagger A_k - A_k^\dagger A_k \rho (t) \right].
\end{equation}
In the above, $\rho (t)$ is the density matrix of the system's quantum state at
time $t$, $H(t)$ is its time-dependent Hamiltonian, and $A_k$ are the operators 
through which the environment couples to the system with rate $\gamma _k$.
As \Qibolab currently only supports superconducting-qubits-based hardware, 
the device model used by the emulator is based on $N$ capacitively coupled 
transmon qubits modelled as Duffing oscillators~\cite{Krantz_2019,Magesan2020} 
with number of energy levels predefined by the user in the \texttt{PulseSimulator} settings:
\begin{eqnarray}
  H (t) &=& H_{\rm sys} + H_{\rm drive} (t), \\
  H_{\rm sys} &=& \sum _{i=1}^N \left( \omega _i b_i ^\dagger b_i + \frac{\alpha_i}{2} b_i ^\dagger b_i (b_i ^\dagger b_i - 1) \right) + \sum _{i \neq j} g_{ij} (b_i ^\dagger b_j + b_i ^\dagger b_j ),\\
  H_{\rm drive} (t) &=& \sum _{i=1}^N \left[\Omega_{X,i} (t) \cos (\omega _{{\rm drive}, i} t) + \Omega_{Y,i} (t) \sin (\omega _{{\rm drive}, i} t) \right] (b_i ^\dagger + b_i).
\end{eqnarray}
In the above, $\omega _i, \omega _{{\rm drive}, i}$ and $\alpha _i$ denote the 
resonant frequency, drive frequency and anharmonicity respectively for transmon
$i$, $b_i (b_i^\dagger)$ its annihilation(creation) operator, and
$\Omega_{X,i}(t), \Omega_{Y,i}(t) $ the drive amplitudes on its quadratures, 
while $g_{ij}$ denotes the coupling strength between transmon $i$ and $j$. 
Decoherence is incorporated using the Bloch-Redfield model~\cite{Krantz_2019},
whereby each transmon is characterized by its longitudinal and transverse 
relaxation times $T_{i,1}$ and $T_{i,2}$ respectively,
\begin{equation}
A_{i,1} = \frac{1}{2}(\sigma _{i,X} + i \sigma_{i,Y}), \quad A_{i,2} = \sigma _{i,Z}, \quad \gamma _{i,1(2)} = \frac{2\pi}{T_{i,1(2)}},
\end{equation}
with $\sigma _{i,\mu = X,Y,Z}$ the Pauli matrices corresponding to transmon $i$.
As an example, we show in Fig.~\ref{fig:sim_plot} the state overlap between the
transmon qubit, modelled as a three-level system, with each of its energy modes
as it interacts with the pulse for the X gate followed by the Hadamard gate.
This can be easily extended to include other quantum computing technologies when
they are available on \Qibolab.
The first supported simulation engine is based on \texttt{QuTiP}~\cite{Qutip},
with plans to incorporate \texttt{JAX}~\cite{jax2018github} support for GPU acceleration, as well as
tensor-network based quantum dynamics simulation libraries to speed up the
simulation of larger system sizes with a lower memory footprint but with a small accuracy cost.

For the initial release, resonators are not included in the simulation model and
therefore the only acquisition modes supported are the discrimination and integration, where the 
latter is currently implemented as a projection onto the in-phase pulse component. 
Despite this limitation, the available emulator can already execute most \Qibocal
protocols to simulate calibration and device characterization.
This part of the library is still a work in progress and we plan to expand it
with new features.  Increasing the complexity of the simulated quantum system with
the inclusion of flux tunable qubits, resonators and couplers is part of our roadmap.

\section{Conclusions}

In this proceedings, we have presented significant updates to \Qibolab, a software library
designed to harness the full potential of \Qibo~for executing quantum algorithms
on self-hosted quantum hardware platforms. The enhancements introduced in \Qibolab
include a more platform-agnostic approach to pulse definitions, streamlined 
orchestration of quantum hardware control through the \texttt{Platform} object,
and the integration of an emulator for simulating quantum hardware behaviour.

Our work goal is a platform-agnostic interface to facilitate
the transition from theoretical quantum computing models to practical experiments.
This approach not only reduces the expertise required to operate diverse quantum
platforms but it also ensures tools reusability as well as experiment and data 
sharing across a community of quantum laboratories. 
This collaborative environment significantly lower the barriers associated with
different experimental setups.

Benchmark results demonstrate that \Qibolab improvements lead to more
efficient execution times for qubit calibration routines, highlighting the 
library's capability to optimize the performance of various quantum control electronics.
The refined boundaries between the \texttt{Platform} and \texttt{Channel} classes
and the relocation of the \texttt{Transpiler} to \Qibo~further streamline the 
software architecture, ensuring a clearer separation of roles and greater transparency.

The addition of the emulator component represents a crucial tool for both software 
testing and predictive analysis of calibration experiments. While currently, it is 
still a work in progress, the emulator sets the stage for future expansions,
including more complex quantum systems and advanced acquisition modes.

In conclusion, the advancements in \Qibolab~not only enhance its functionality 
and efficiency but also contribute to the broader goal of creating an extensible,
quantum hardware-agnostic, open-source hybrid quantum operating system. This system,
fully tested and benchmarked on superconducting platforms, offers a robust foundation
for ongoing research and development in the field of quantum computing.
By fostering a collaborative community and providing powerful, flexible tools,
\Qibolab~and \Qibo~collectively advance the frontier of quantum technology, 
supporting both current and future innovations in quantum research and applications.

\section{Acknowledgments}
This project is supported by \texttt{TII}’s Quantum Research Center. 
This work was further supported by the National Research Foundation Singapore, under its Quantum Engineering Programme 2.0 (National Quantum Computing Hub, NRF2021-QEP2-02-P01).
The authors thank all \Qibo contributors for helpful discussion and Dr. Ye Jun. 
M.R. is supported by \texttt{CERN}’s Quantum Technology Initiative (\texttt{QTI}) through the Doctoral Student Program.
J.Y.K acknowledges funding support from A*STAR C230917003.

\bibliographystyle{unsrtnat}
\bibliography{references}

\begin{thebibliography}{13}
\providecommand{\natexlab}[1]{#1}
\providecommand{\url}[1]{\texttt{#1}}
\expandafter\ifx\csname urlstyle\endcsname\relax
  \providecommand{\doi}[1]{doi: #1}\else
  \providecommand{\doi}{doi: \begingroup \urlstyle{rm}\Url}\fi

\bibitem[Efthymiou et~al.(2024)Efthymiou, Orgaz-Fuertes, Carobene, Cereijo,
  Pasquale, Ramos-Calderer, Bordoni, Fuentes-Ruiz, Candido, Pedicillo,
  Robbiati, Tan, Wilkens, Roth, Latorre, and Carrazza]{Efthymiou_2024}
Stavros Efthymiou, Alvaro Orgaz-Fuertes, Rodolfo Carobene, Juan Cereijo, Andrea
  Pasquale, Sergi Ramos-Calderer, Simone Bordoni, David Fuentes-Ruiz,
  Alessandro Candido, Edoardo Pedicillo, Matteo Robbiati, Yuanzheng~Paul Tan,
  Jadwiga Wilkens, Ingo Roth, José~Ignacio Latorre, and Stefano Carrazza.
\newblock Qibolab: an open-source hybrid quantum operating system.
\newblock \emph{Quantum}, 8:\penalty0 1247, February 2024.
\newblock ISSN 2521-327X.
\newblock \doi{10.22331/q-2024-02-12-1247}.
\newblock URL \url{http://dx.doi.org/10.22331/q-2024-02-12-1247}.

\bibitem[Efthymiou et~al.(2021)Efthymiou, Ramos-Calderer, Bravo-Prieto,
  P{\'{e}}rez-Salinas, Garc{\'{\i}}a-Mart{\'{\i}}n, Garcia-Saez, Latorre, and
  Carrazza]{Efthymiou_2021}
Stavros Efthymiou, Sergi Ramos-Calderer, Carlos Bravo-Prieto, Adri{\'{a} }n
  P{\'{e}}rez-Salinas, Diego Garc{\'{\i}}a-Mart{\'{\i}}n, Artur Garcia-Saez,
  Jos{\'{e}}~Ignacio Latorre, and Stefano Carrazza.
\newblock Qibo: a framework for quantum simulation with hardware acceleration.
\newblock \emph{Quantum Science and Technology}, 7\penalty0 (1):\penalty0
  015018, dec 2021.
\newblock \doi{10.1088/2058-9565/ac39f5}.

\bibitem[Pasquale et~al.(2024)Pasquale, Efthymiou, Ramos-Calderer, Wilkens,
  Roth, and Carrazza]{Qibocal}
Andrea Pasquale, Stavros Efthymiou, Sergi Ramos-Calderer, Jadwiga Wilkens, Ingo
  Roth, and Stefano Carrazza.
\newblock Towards an open-source framework to perform quantum calibration and
  characterization, 2024.
\newblock URL \url{https://arxiv.org/abs/2303.10397}.

\bibitem[ZurichInstruments(2023)]{LabOneQ}
ZurichInstruments.
\newblock \url{
  https://www.zhinst.com/others/en/quantum-computing-systems/labone-q }, 2023.

\bibitem[Qblox()]{Qblox}
Qblox.
\newblock \url{https://www.qblox.com}.

\bibitem[Ella et~al.(2023)Ella, Leandro, Wertheim, Romach, Szmuk, Knol, Ofek,
  Sivan, and Cohen]{qm_qua}
Lior Ella, Lorenzo Leandro, Oded Wertheim, Yoav Romach, Ramon Szmuk, Yoel Knol,
  Nissim Ofek, Itamar Sivan, and Yonatan Cohen.
\newblock Quantum-classical processing and benchmarking at the pulse-level,
  2023.
\newblock URL \url{https://doi.org/10.48550/arXiv.2303.03816}.

\bibitem[Stefanazzi et~al.(2022)Stefanazzi, Treptow, Wilcer, Stoughton,
  Bradford, Uemura, Zorzetti, Montella, Cancelo, Sussman, Houck, Saxena,
  Arnaldi, Agrawal, Zhang, Ding, and Schuster]{Stefanazzi2022}
Leandro Stefanazzi, Kenneth Treptow, Neal Wilcer, Chris Stoughton, Collin
  Bradford, Sho Uemura, Silvia Zorzetti, Salvatore Montella, Gustavo Cancelo,
  Sara Sussman, Andrew Houck, Shefali Saxena, Horacio Arnaldi, Ankur Agrawal,
  Helin Zhang, Chunyang Ding, and David~I. Schuster.
\newblock The {QICK} (quantum instrumentation control kit): Readout and control
  for qubits and detectors.
\newblock \emph{Review of Scientific Instruments}, 93\penalty0 (4), April 2022.
\newblock \doi{10.1063/5.0076249}.
\newblock URL \url{https://doi.org/10.1063/5.0076249}.

\bibitem[Carobene et~al.(2023)Carobene, Candido, Serrano, Orgaz-Fuertes,
  Giachero, and Carrazza]{Qibosoq}
Rodolfo Carobene, Alessandro Candido, Javier Serrano, Alvaro Orgaz-Fuertes,
  Andrea Giachero, and Stefano Carrazza.
\newblock Qibosoq: an open-source framework for quantum circuit rfsoc
  programming, 2023.
\newblock URL \url{https://arxiv.org/abs/2310.05851}.

\bibitem[Qcodes(2023)]{QCODESsite}
Qcodes.
\newblock \url{https://qcodes.github.io/Qcodes/}, 2023.

\bibitem[Krantz et~al.(2019)Krantz, Kjaergaard, Yan, Orlando, Gustavsson, and
  Oliver]{Krantz_2019}
P.~Krantz, M.~Kjaergaard, F.~Yan, T.~P. Orlando, S.~Gustavsson, and W.~D.
  Oliver.
\newblock {A quantum engineer's guide to superconducting qubits}.
\newblock \emph{Applied Physics Reviews}, 6\penalty0 (2):\penalty0 021318, 06
  2019.
\newblock ISSN 1931-9401.
\newblock \doi{10.1063/1.5089550}.
\newblock URL \url{https://doi.org/10.1063/1.5089550}.

\bibitem[Magesan and Gambetta(2020)]{Magesan2020}
Easwar Magesan and Jay~M. Gambetta.
\newblock Effective hamiltonian models of the cross-resonance gate.
\newblock \emph{Phys. Rev. A}, 101:\penalty0 052308, May 2020.
\newblock \doi{10.1103/PhysRevA.101.052308}.
\newblock URL \url{https://link.aps.org/doi/10.1103/PhysRevA.101.052308}.

\bibitem[Johansson et~al.(2013)Johansson, Nation, and Nori]{Qutip}
J.R. Johansson, P.D. Nation, and Franco Nori.
\newblock Qutip 2: A python framework for the dynamics of open quantum systems.
\newblock \emph{Computer Physics Communications}, 184\penalty0 (4):\penalty0
  1234--1240, 2013.
\newblock ISSN 0010-4655.
\newblock \doi{https://doi.org/10.1016/j.cpc.2012.11.019}.
\newblock URL
  \url{https://www.sciencedirect.com/science/article/pii/S0010465512003955}.

\bibitem[Bradbury et~al.(2018)Bradbury, Frostig, Hawkins, Johnson, Leary,
  Maclaurin, Necula, Paszke, Vander{P}las, Wanderman-{M}ilne, and
  Zhang]{jax2018github}
James Bradbury, Roy Frostig, Peter Hawkins, Matthew~James Johnson, Chris Leary,
  Dougal Maclaurin, George Necula, Adam Paszke, Jake Vander{P}las, Skye
  Wanderman-{M}ilne, and Qiao Zhang.
\newblock {JAX}: composable transformations of {P}ython+{N}um{P}y programs,
  2018.
\newblock URL \url{http://github.com/google/jax}.

\end{thebibliography}
\end{document}